\documentclass{aa}

\usepackage{natbib}
\usepackage{longtable}
\usepackage{lscape}
\usepackage{amsmath,amssymb}
\usepackage{txfonts}
\usepackage{color}
\usepackage{url}
\usepackage{xspace}
\usepackage{graphicx}

\usepackage{ulem}

\bibpunct{(}{)}{;}{a}{}{,}

\newcommand{\kev}{keV\xspace}
\newcommand{\xmm}{\textsl{XMM-Newton}\xspace}
\newcommand{\rosat}{\textsl{ROSAT}\xspace}
\newcommand{\swift}{\textsl{SWIFT}\xspace}

\def\aplt{\ {\raise-.5ex\hbox{$\buildrel<\over\sim$}}\ }

\begin{document}
   \title{A 33\,hour period for the Wolf-Rayet/black hole X-ray~binary candidate NGC~300~X-1}

   \author{S. Carpano
          \inst{1}
          \and
	  A.~M.~T. Pollock
          \inst{1}
          \and
	  A. Prestwich 
	  \inst{2}
	  \and
	  P. Crowther
	  \inst{3}
	  \and
          J. Wilms
          \inst{4}
	  \and
	  L. Yungelson
	  \inst{5}
	  \and
	  M. Ehle
	  \inst{1}}

   \offprints{S. Carpano, e-mail: scarpano@sciops.esa.int}
   \institute{XMM-Newton Science Operations Centre, ESAC, ESA, PO Box 50727, 28080 Madrid, Spain
             \and
	     Harvard-Smithsonian Center for Astrophysics, Cambridge, MA 02138, USA
	     \and
	     Department of Physics \& Astronomy, University of Sheffield, Hicks Building, Hounsfield Rd, Sheffield S3 7RH, UK
	     \and
	     Dr. Remeis-Observatory, Astronomisches Institut der FAU Erlangen-N\"urnberg, Sternwartstr. 7, 96049 Bamberg, Germany
	     \and
	     Institute of Astronomy of the Russian Academy of Sciences, 48 Pyatnitskaya Str., 119017 Moscow, Russia 
              }

   \date{Submitted: 27 February 2007; Accepted: 10 March 2007}
   
   \abstract{NGC~300~X-1 is the second extragalactic candidate, after IC~10~X-1, in the rare class of 
   Wolf-Rayet/compact object X-ray binary systems exemplified in the Galaxy by Cyg~X-3.
   From a theoretical point of view, accretion 
    onto a black hole in a detached system is possible for large
   orbital periods only if the mass of the relativistic object is high or the velocity of the accreted wind is low.}
   {We analysed a 2 week  \swift XRT light curve of NGC~300~X-1 and searched for periodicities.}
   {Period searches were made using  Lomb-Scargle periodogram analysis. We evaluated the confidence level 
   using Monte Carlo simulations.}{A period of 32.8$\pm$0.4\,h (3$\sigma$ error) 
   was found for NGC~300~X-1 with a confidence level $>$99\%. Furthermore, we confirm the high irregular variability 
   during the high flux level, as
   already observed in the \xmm observations of the source. A folded \xmm light curve is shown, with a
   profile that is in agreement with \swift. The mean absorbed X-ray luminosity in
   the \swift observations was
   $1.5\times10^{38}$\,erg~s$^{-1}$, close to the value derived from the \xmm data.}
   {While Cyg~X-3 has a short period of 4.8 h, the period of NGC~300~X-1 is very close to that of IC~10~X-1 
    (34.8$\pm$0.9\,h). These are likely orbital periods. Possibility of formation of 
    accretion disk for such high orbital periods  strongly depends on the terminal velocity of the Wolf-Rayet
   star wind and black-hole mass. While low masses are possible for wind velocities $\lesssim$ 1000 km s$^{-1}$, 
   these increase to several tens of solar masses 
   for velocities $>$ 1600 km s$^{-1}$ and no accretion disk may form for terminal velocities larger 
   than 1900 km s$^{-1}$.}

     \keywords{X-rays: individual: NGC~300~X-1 -- X-rays: binaries  -- Stars: Wolf-Rayet} 

\maketitle
%
%____________________________________________________________________________

\section{Introduction}
\label{sec:int}

Wolf-Rayet/black hole binaries are believed to be 
 stars in the evolutionary stage following high-mass X-ray binaries.
The existence of  helium-star/compact object X-ray binaries was
suggested  independently by \cite{vandenHeuvel1973}  to explain the
nature of the galactic source \object{Cyg X-3}, and by \citet{TY73} based on results of evolutionary computations. 
To appear  as an X-ray source, an accretion disk must form and hence the velocity of the Wolf-Rayet (WR) star wind
must be slow enough for the  material around the compact object to be accreted. According to \cite{Illarionov1975},
a black hole appears as a strong X-ray source in a detached binary system only when the orbital period,
$P_{\text{orb}}$:
\begin{equation}
P_{\text{orb}}\aplt4.8\frac{M_{\text{BH}}}{v^4_{1000} \delta^2}\,\text{(h)}
\label{equ:period}
\end{equation}
where $M_{\text{BH}}$ is the black hole mass is solar units,
 $v_{1000}$ is the velocity of the accreted wind in units of 1000\,km\,s$^{-1}$ and $\delta$ is a dimensionless parameter
 of order unity.  It has been shown by \cite{Ergma1998} and \cite{Lommen2005} that these periods
  for solar metallicity stars cannot be larger than several tens of hours.

So far, Cyg~X-3 is the only valid candidate in our galaxy for a Wolf-Rayet/compact object X-ray binary system.
Its X-ray luminosity is high, $L_\text{X}\sim10^{38}$\,erg~s$^{-1}$. 
The companion star was identified as a WR star by \cite{vanKerkwijk1992} and
then designated as WR~145a in the 7th catalogue of galactic 
Wolf-Rayet stars \citep{vanderHucht2001}.
Its orbital period is very short, 4.8\,h \citep{Parsignault1972}.

\object{IC~10~X-1} ($L_\text{X}\sim1.2\times10^{38}$\,erg~s$^{-1}$), in the starburst galaxy \object{IC~10} 
located at 0.8\,Mpc, was the first extragalactic candidate for this class of objects \citep{Bauer2004,Wang2005}.
A period of 34.8$\pm$0.9\,h has been observed recently thanks to \swift observations 
(A. Prestwich et al. , ATel \#955,  paper in preparation). We report in this Letter the discovery of 
a very similar but slightly shorter period of 32.8\,h for NGC~300~X-1,
which is the second extragalactic Wolf-Rayet/compact object X-ray binary candidate \citep{Carpano2007}.

\object{NGC~300~X-1} is the brightest X-ray point source in the dwarf spiral galaxy \object{NGC 300}
at a distance of $\sim$1.88\,Mpc \citep{Gieren2005}. The galaxy is almost face-on
and has a low Galactic column density of $N_\text{H}=3.6\times10^{20}\,\text{cm}^{-2}$ \citep{Dickey1990}.
Study of its X-ray population has been done by \cite{Read2001} using \rosat and by 
\cite{Carpano2005} using \xmm.
Based on the existing four \xmm observations, it has been shown
 in \cite{Carpano2007} that the position 
of the X-ray source ($\alpha_\text{J2000}=00^\text{h}55^\text{m} 10\fs{}00$,
$\delta_\text{J2000}=-37^\circ 42' 12\farcs 06$) coincides with a WR candidate, WR~41 \citep{Schild2003}, 
within  $0\farcs11\pm0\farcs45$. WR~41 has now been spectroscopically confirmed as an early-type WN star
 (Crowther et al., in preparation).

The four \xmm light curves, lasting $\sim$10\,h each, showed irregular variability, and during one 
observation, the flux increased by about a factor of ten in 10\,h. No period  between 
5\,sec and 30\,ksec (8.3\,h) was 
found in the data. The mean observed (absorbed) luminosity in the 0.2--10\,\kev band was
$\sim2\times10^{38}\,\text{erg}\,\text{s}^{-1}$. The unabsorbed X-ray luminosity reached
L$_{0.2-10\,\text{\kev}}\sim1\times10^{39}\,\text{erg}\,\text{s}^{-1}$ suggesting the presence 
of a black hole, altough beamed emission from a neutron star cannot be excluded.
 The spectrum could be modelled by a power-law with $\Gamma\sim2.45$ with additional 
relatively weak  emission, notably around 0.95\,\kev.

In this Letter, we report the discovery of a 32.8\,h period for NGC~300~X-1. The remainder of the 
Letter is organised as follows.
Section~\ref{sec:obs} briefly describes the  \swift
observations and data reduction. In Sect.~\ref{sec:time}, we report
analysis of the \swift XRT light curve and search for periodicities using a 
Lomb-Scargle periodogram analysis.
A folded \xmm light curve is shown in  
Sect.~\ref{sec:xmm}, while a discussion of our results is given in
Sect.~\ref{sec:conc}.

\section{Observations and data reduction}
\label{sec:obs}
NGC~300~X-1 was observed with the \swift Gamma-Ray Burst Explorer \citep{Gehrels2004}
between 2006 December 26 and 2007 January 10, for a total of 83\,ksec. The light curve
of NGC~300~X-1 was extracted from the X-ray Telescope, XRT \citep{Burrows2005}
which operates in the 0.2--10\,\kev energy band.
There are 146 XRT observations lasting between 10 and 1477\,sec. We kept 
only data from 124 observations lasting more than 100\,sec.

For the production of the X-ray light curve, we analysed the calibrated and screened 
PC event files (level~2) provided in the set of data products. Source and background regions were
extracted using the \texttt{FTOOLS}\footnote{\url{http://heasarc.nasa.gov/lheasoft/ftools}} 
\texttt{ftselect} task. The circular source region was centred 
on the \xmm source position ($\alpha_\text{J2000}=00^\text{h}55^\text{m} 10\fs{}00$,
$\delta_\text{J2000}=-37^\circ 42' 12\farcs 06$) with a radius of $40''$, which is larger
 than the telescope PSF 
($18''$). A similar sized region was extracted for the background, in a blank region close to NGC~300~X-1.

%_____________________________________________________________________________

\section{Time analysis of \swift light curve}
\label{sec:time}
\begin{figure}
  \resizebox{\hsize}{!}{\includegraphics{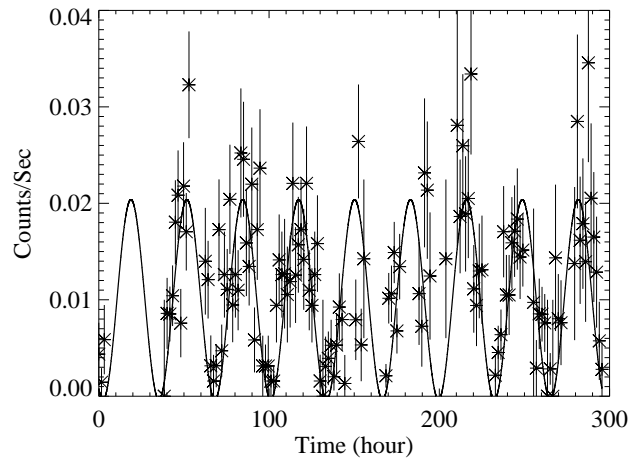}}
 \caption{\swift XRT background-substracted light curve of NGC~300~X-1. 
 Overplotted is the best-fit sinusoid function.}
 \label{fig:swiftlc}
\end{figure}

The \swift XRT background-subtracted light curve is shown in Fig.~\ref{fig:swiftlc}. Times are
given in hours from the beginning of the observation. We overplotted the best-fit sinusoid function.
For clarity, the amplitude has been multiplied by a factor of 1.5. 
It is clear that the flux varies in a regular way, with the minima likely to be eclipses
of the accreting companion.  

\begin{figure}
  \resizebox{\hsize}{!}{\includegraphics{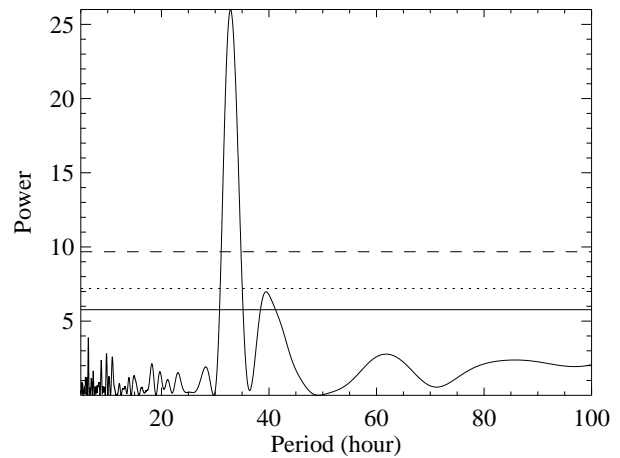}}
 \caption{Search for periodicities for the \swift XRT light curve of NGC~300~X-1 using a
 Lomb-Scargle periodogram analysis. The full, dotted and dashed lines represent the 68\%, 90\% and
 99\% confidence level respectively.}
 \label{fig:period}
\end{figure}

\begin{figure}
  \resizebox{\hsize}{!}{\includegraphics{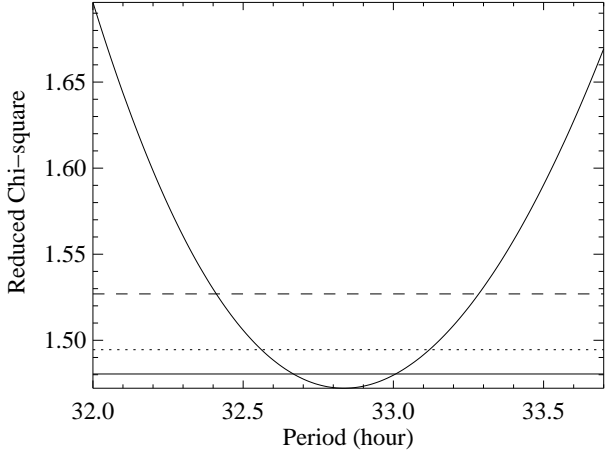}}
 \caption{Period error search fitting sine curve. The full, dotted and dashed lines represent 
 $\Delta \chi^2=$1.00, 2.71 and 6.63 respectively. }
 \label{fig:perioder}
\end{figure}

We searched for a periodic signal between 5 and 100\,h, using a Lomb-Scargle periodogram analysis
\citep{Lomb1976,Scargle1982}. By means of  Monte Carlo simulations, we evaluated 
the confidence level assuming a null hypothesis of
white noise. Results are plotted in Fig.~\ref{fig:period}. The full, dashed 
and dotted lines represent the 68\%, 90\% and 99\% confidence level respectively. 
 We found that the
32.84\,h period is significant at a confidence level $>99\%$. 
To estimate the error, we fitted a sine function using the IDL task \texttt{curvefit} 
keeping trial periods fixed. The reduced chi-square, $\chi^2_\nu$ with $\nu$=121, is
 shown in Fig.~\ref{fig:perioder}. The full, dotted 
and dashed lines represent $\Delta \chi^2=$1.00, 2.71 and 6.63 respectively. 
The corresponding 1, 2 and 3 $\sigma$ period range are
$[32.67-33.00]$, $[32.57-33.12]$ and $[32.42-33.28]$ respectively.
Note that the $\chi^2_\nu$  larger than 1 shows that the light curve cannot be described
by a pure sinusoid function.

\begin{figure}
  \resizebox{\hsize}{!}{\includegraphics{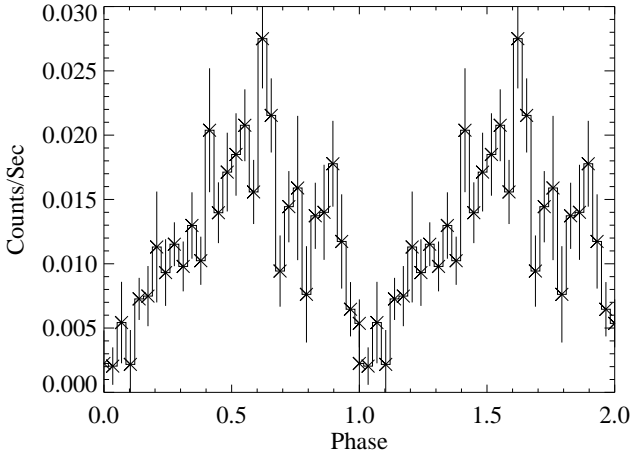}}
 \caption{\swift XRT light curve folded at 32.84\,h using 30 bins. 
 Phase zero is associated to the 
 beginning of the first \swift  observation.}
 \label{fig:fold}
\end{figure}

The \swift XRT light curve folded at 32.84\,h is shown in  Fig.~\ref{fig:fold}.
 Phase zero is associated to the beginning of the first \swift  observation.
From Fig.~\ref{fig:swiftlc} and Fig.~\ref{fig:fold}, we confirm irregular high variability outside the
eclipse as observed in the \xmm data \citep{Carpano2007}.

We estimate the X-ray luminosity by converting the mean count rate, 0.012 count\,s$^{-1}$ to flux with
\texttt{WebPIMMS}\footnote{\url{http://heasarc.gsfc.nasa.gov/Tools/w3pimms.html}}, using the spectral
parameters derived by \cite{Carpano2007}. The mean absorbed luminosity in the 0.2--10\,\kev energy band
is $1.5\times10^{38}$erg\,s$^{-1}$, which is close to that found from the \xmm data, 
$L_{0.2-10\,\text{\kev}}\sim2\times10^{38}\,\text{erg}\,\text{s}^{-1}$ \citep{Carpano2007} and close to the ROSAT value,
$L_{0.1-2.4\,\text{\kev}}\sim2.2\times10^{38}\,\text{erg}\,\text{s}^{-1}$ \citep{Read2001}.

%_____________________________________________________________________________

\section{\xmm folded light curve}
\label{sec:xmm}
We searched for  periodicity in the four \xmm data  samples
observed between 2000 and 2005 (see \cite{Carpano2006} 
for more details about the observations). Although the unfavourable sampling precludes a 
rigorous period search, we have used some \textit{a priori} information we have on the eclipse profile to fold the light
curve. From Table.~1 and Fig.~2 of \cite{Carpano2007}, we can compare 
the flux and light curve shape of the several observations to the \swift
profile of Fig.~\ref{fig:fold}. 
The flux was minimum in the first \xmm observation and began to increase at the end. 
This could be associated to the eclipse state of the X-ray source. The second
\xmm observation, 6~days later, is likely associated with an eclipse egress. In the third
observation the flux is lower than in the second and fourth observations, and shows a small
decrease trend: this data set could be associated with the beginning of the eclipse 
ingress. And in the fourth \xmm observation, the flux is  high and likely outside the
eclipse. We used this information to constrain the phase of each beginning of data set, in the folded
light curve. Only few values of the period around 33\,h, but larger than 32.8\,h, are possible to 
provide a reasonable profile.

\begin{figure}
  \resizebox{\hsize}{!}{\includegraphics{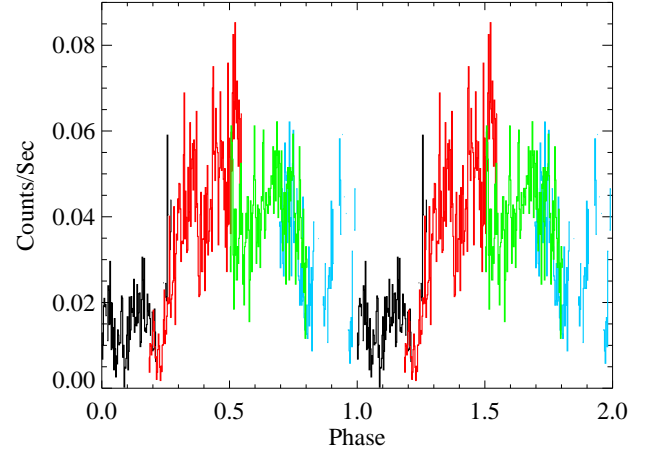}}
 \caption{MOS \xmm light curve folded at a period of 33.066\,h using time bins of 300\,sec \citep{Carpano2007}.
 First observation is in black, second in red, third in blue and fourth in green.
 Phase zero is associated with the start of the \xmm observations and is by chance during 
eclipse as for the \swift data.}
 \label{fig:xmmfold}
\end{figure}

 Fig.~\ref{fig:xmmfold} shows the \xmm EPIC MOS light curve folded at a period of 33.066\,h, 
  within the 2\,$\sigma$ error of the \swift period. 
Phase zero is associated with the start of the \xmm observations and is by chance during 
eclipse as for the \swift data. Comparing Fig.~\ref{fig:fold} and Fig.~\ref{fig:xmmfold},
it seems that both curves have a small dip around phase 0.6--0.8. Further observations
of the source are clearly necessary to confirm this and other features of the periodic light curve.

%\begin{figure}
%  \resizebox{\hsize}{!}{}
% \caption{}
% \label{fig:}
%\end{figure}

%_____________________________________________________________________________

\section{Discussion}
\label{sec:conc}

In the evolutionary scenario for WR/compact object X-ray binaries that has
been suggested to explain the short orbital period observed in Cyg~X-3,
the immediate precursor of the system is a neutron star or black hole 
orbiting  an OB star. When the latter leaves the main sequence, matter is transferred
due to Roche lobe overflow and a common envelope forms. Due to friction, the distance between the
early-type star core and the compact object decreases. A merger is avoided
if the binding energy of the hydrogen envelope is lower than the energy released by spiral-in,
leading to a short-period binary system consisting of a WR star and a compact object.
Formation of an accretion disk around a black hole from the strong wind of the helium star is then possible
 if the orbital period satisfies  Eq.~(\ref{equ:period}). 

\begin{figure}
  \resizebox{\hsize}{!}{\includegraphics{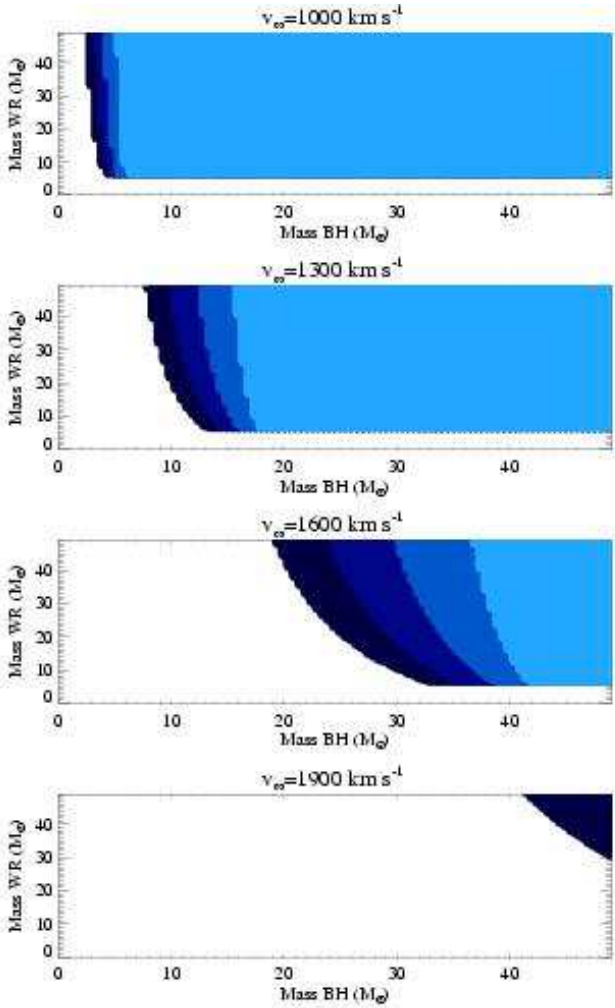}}
 \caption{Allowed WR and BH masses, in solar  units, that satisfy the condition for the formation of an accretion 
 disk for different values of the terminal velocity of the wind.
 Different values of the wind velocity $\beta$ parameter (2.0, 1.5, 1.0 and 0.5) are represented by dark to 
 light blue. For each value of $\beta$, the allowed region of M$_\text{WR}$--M$_\text{BH}$ parameter space extents
 all the way up to the right. }
 \label{fig:wind}
\end{figure}

We now derive  possible values for the masses of the black hole 
and the Wolf-Rayet star that allow the formation of an accretion disk for an
orbital period of 32.8\,h.
Kepler's third law gives:
\begin{equation}
a=0.506 P_{\text{orb}}^{2/3} (M_\text{BH}+M_\text{WR})^{1/3}\,\,(\text{R}_\odot)
\label{equ:kepler}
\end{equation}
where, for a circular orbit, $a$ is the binary separation, $P_{\text{orb}}$
the orbital period in hours, and $M_\text{BH}$ and $M_\text{WR}$ the component masses in solar units.
The velocity of the Wolf-Rayet wind, $v_\text{WR}$ at  $a$ 
can be approximated by a $\beta$-law
\citep{Lamers1999}:
\begin{equation}
v_\text{WR}(a)=v_0+(v_\infty-v_0)\left(1-\frac{R_\text{WR}}{a}\right)^\beta
\label{equ:lammers}
\end{equation}
where $R_\text{WR}$ is the Wolf-Rayet star radius (in  $R_\odot$), $v_\infty$ the terminal velocity,
 $v_0$ the initial velocity ($\sim$0.01$v_\infty$) and the $\beta$ parameter describes 
 the steepness of the law. Note that $\beta=1$ is the preferred value for Wolf-Rayet winds \citep{Grafener2005}.
 For the radii of Wolf-Rayet stars, we can use the relation given by 
\cite{Schaerer1992}:
\begin{equation}
\log(R_\text{WR})=-0.6629+0.5840 \log(M_\text{WR}).
\label{equ:schaerer}
\end{equation}
 Given that orbital period of the system is known, Eqs.~(\ref{equ:period}) -- (\ref{equ:schaerer}) define 
combinations of the component masses
 for which an accretion disk may form. The plots are shown in Fig.~\ref{fig:wind} for different values of the
$\beta$ parameter  and  the terminal velocity.
We restrict to M$_\text{WR}>$7\,M$_\odot$ and M$_\text{BH}>$3\,M$_\odot$, which are standard lower 
limit for the WR and BH masses respectively.
 
Looking at these graphs we can note that, for a terminal velocity around 1000\,km s$^{-1}$, whatever the mass
of the Wolf-Rayet star and the value of the $\beta$ parameter, the lower limit for the  black hole mass
is  below  7\,M$_\odot$. With higher values for the terminal velocity, this lower limit 
increases significantly and becomes more and more dependent on the $\beta$ parameter. For velocities of
1600 km s$^{-1}$, the mass must be at least  several tens of solar masses, while no accretion disk 
may form for terminal velocities significantly higher than 1900 km s$^{-1}$. 
From the optical spectrum of WR~41, the terminal velocity of the wind is about 1250\,km\,s$^{-1}$
and the mass of the Wolf-Rayet star is estimated between 18\,M$_\odot$ and 40\,M$_\odot$
(Crowther et al., in preparation). This leads to a black hole mass for NGC~300~X-1 larger than 13\,M$_\odot$
for $\beta=1$.

Similar arguments apply for IC~10~X-1: the mass of the Wolf-Rayet star was derived to be 35\,M$_{\odot}$ and
its  terminal wind   velocity $1750$ km s$^{-1}$ \citep{Clark2004}. This means that the mass 
of the black hole companion must be at least of $\sim$35\,M$_{\odot}$, for an orbital period of 34.8\,h.

To conclude, it seems a surprise that orbital periods found in both extragalactic WR/compact object X-ray binary
candidates IC~10~X-1 and NGC~300~X-1 are so similar. Furthermore,  their difference to the short period of Cyg X-3 
may suggest different paths of evolution.

\begin{acknowledgements}
This paper is based on observations obtained with the \swift gamma-ray burst mission
and observations from \textsl{XMM-Newton}, an ESA
  science mission with instruments and contributions directly finded
  by ESA Member States and NASA.  LY is supported by the Russian Academy of Sciences grant 
 ``Origin and evolution of stars and galaxies'' and NSF grant
No. PHY99-07949.
We warmly  thank Neil Gehrels and Dave Burrows for approving the SWIFT observing time.
\end{acknowledgements}

\end{document}